# ODMR Of Impurity Centers Embedded In Silicon Microcavities

N.T. Bagraev[a], W. Gehlhoff[b], L.E. Klyachkin[a], A.M. Malyarenko[a], V.A. Mashkov[a], V.V. Romanov[a] and T.N. Shelykh[a]

[a]Ioffe Physico-Technical Institute, 194021, St.Petersburg, Russia
[b]Technische Universitaet Berlin, D-10623, Berlin, Germany

**Abstract.** We present the findings of high efficient light absorption in self-assembled quantum wells (SQW) embedded in silicon microcavities that exhibit a distributed feedback identified by the FIR transmission spectra. The excitonic normal-mode coupling (NMC) is found to result in high efficient bound exciton photoluminescence in the range of the Rabi splitting. The bound excitons at the iron-boron pair and the erbium-related centers inserted in SQW are shown to cause giant exchange splitting of the center multiplets as a result of the strong *sp-d* and *sp-f* mixing in the absence of the external magnetic field. The NMC regime is observed to reveal this exchange splitting in the angular dependencies of the transmission spectra measured in the range of the Rabi splitting that are evidence of the ODMR of the trigonal iron-boron pairs and trigonal erbium-related centers.

## INTRODUCTION

The diffusion of boron is known to be controlled by means of adjusting the fluxes of self-interstitials and vacancies thereby forming the self-assembled silicon quantum wells (SQW) of the *p*-type confined by the heavily doped delta barriers on the *n*-type Si (100) surface (Fig. 1a) [1]. The boron centers inside the delta barriers are found to be the impurity dipoles, $B^+-B^-$, which cause the GHz generation under applied voltage [2]. Spectroscopic studies confirmed this pattern and furthermore showed that SQW are incorporated into the silicon microcavities which are created between self-assembled microdefects induced by the same fluxes of self-interstitials [1]. The goal of the present work is to use the exciton normal-mode coupling (NMC) with a single SQW in a 1λ silicon microcavity in studies of single point defects embedded in nanostructures.

## METHODS

The *n*- type Si (100) wafers preliminary doped with iron and erbium were oxidized at 1150°C in dry oxygen containing $CCl_4$ vapors. Short-time boron doping was done from the gas phase under fine surface injection of both self-interstitials and vacancies into the window, which was cut in the oxide overlayer after preparing a mask and performing the subsequent photolithography.

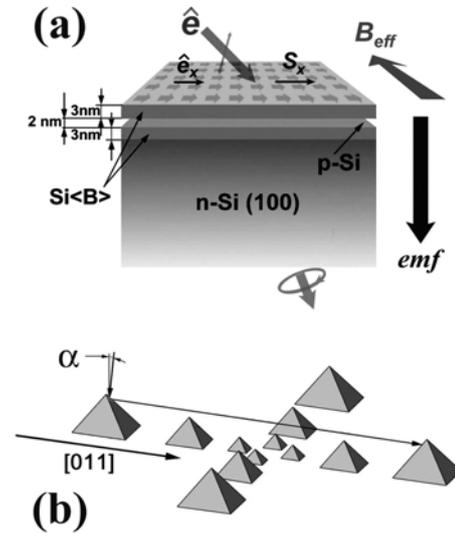

**FIGURE 1.** The ODMR transmission experiment with the silicon quantum well of the *p*-type that is confined by the delta barriers containing the dipole centers of boron on the *n*-type Si (100) surface. The effective magnetic field caused by the Rashba spin-orbit interaction, $B_{eff}$, results from the emf effect generated under optical illumination with linearly polarized light (a). The p-type quantum well doped with transition or rare-earth metals is introduced into the self-assembled microcavity system (b).

## RESULTS

The four-point probe under layer-by-layer etching and SIMS measurements allowed the depth of the $p^+$-diffusion profile, 8 nm. The CR angular dependencies have shown that the profile prepared contains the low density $p$-type SQW confined by the delta barriers heavily doped with boron (Fig. 1a) [1]. Besides, the STM images have demonstrated that this single SQW is incorporated into the microcavity system of the fractal type formed by the microdefects of the self-interstitials type (Fig. 1b).

These silicon microcavities are revealed by the angle-resolved transmission spectra that exhibit the exciton normal-mode coupling (NMC) with a single SQW in the spectral range of the Rabi splitting at T=300 K (Figs. 2a, 2b and 3) [3]. Moreover, the bound excitons at a single point defect appeared to cause giant exchange splitting in the absence of the external magnetic field that is created by strong coupling between $d$ or $f$-electron states of the center and the $s$-$p$ electronic states of the host SQW. This strong $sp$-$d$ or $sp$-$f$ mixing is revealed by the angle-resolved photoluminescence and absorption that seem to be the ODMR spectra in zero magnetic field under the NMC conditions (Figs. 2b and 3), because the EPR frequency is able to be selected from the GHz range generated by SQW being in self-agreement the exchange splitting value. The EPR frequency values, 87 GHz (Fig. 2b) and 200 GHz (Fig. 3), that allow the transmission spectra under the ODMR conditions appeared to correlate with the magnitude of the bound exciton line splitting. The angular dependencies of the peak positions in the transmission spectra observed are found to be in a good agreement with the angular dependencies of the EPR line positions of the trigonal iron-boron pair and trigonal erbium-related center calculated respectively at ν=87 GHz and ν=200 GHz.

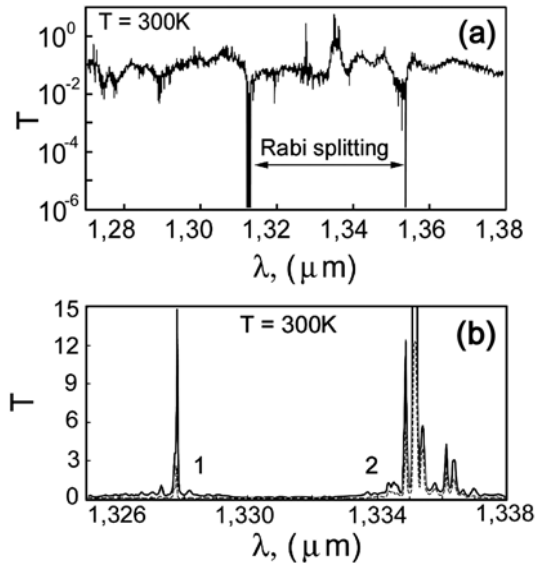

**FIGURE 2.** The spectral dependence of the light transmission coefficient (a) that demonstrates at T=300 K the normal-mode coupling with a single self-assembled silicon quantum well in a 1λ microcavity prepared on the Si(100) surface as well as the σ+ (solid line) and σ- (dashed line) photoluminescence (b) from the self-assembled single silicon quantum well (1) and the trigonal iron-boron pair (2).

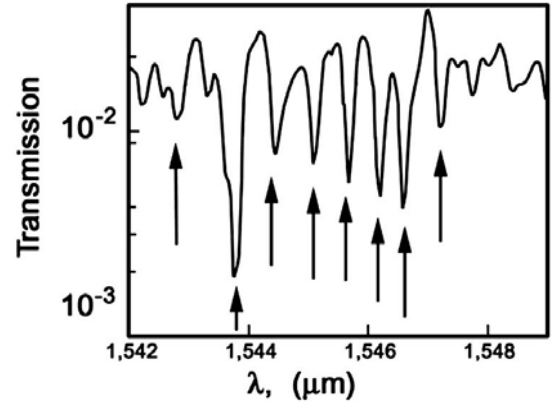

**FIGURE 3.** Transmission spectra that exhibit the $^4I_{13/2}$ => $^4I_{15/2}$ $Er^{3+}$- intracenter multiplet absorption of the trigonal erbium-related center embedded in the self-assembled silicon quantum well microcavity structure, which reveals the hyperfine transitions (J=7/2) in the range of the Rabi splitting. T=300 K.

## SUMMARY

The exciton normal-mode coupling (NMC) with the single QW embedded in the 1λ silicon microcavity on the Si(100) wafer has been identified at T=300 K in the studies of the transmission spectra which revealed the ODMR of the single trigonal iron-boron pair and trigonal erbium-related center in zero magnetic field.

## REFERENCES

1. N.T. Bagraev et al., *Defect and Diffusion Forum* **194-199** 673-678 (2001).
2. N.T. Bagraev et al., *Physica C* **437-438** 21-24 (2006).
3. R. Houdre et al., *Phys.Rev.B* **49** 16761-16764 (1994).